\documentclass[preprint,preprintnumbers,amsmath,amssymb]{revtex4}
\usepackage{graphicx}
\usepackage{bm}

\usepackage{mhchem}
\usepackage{bm}
\usepackage{amsmath}
\usepackage{amsfonts}
\usepackage{amssymb}

\usepackage{breqn}

\usepackage{balance}

\usepackage{times,mathptmx}

\newcommand{\be}{\begin{equation}}
\newcommand{\ee}{\end{equation}}
\newcommand{\bea}{\begin{eqnarray}}
\newcommand{\eea}{\end{eqnarray}}

\begin{document}
\title{The emergence of supramolecular forces from lattice kinetic models of non ideal fluids: applications to the rheology of soft glassy meterials}
\author{M. Sbragaglia$^{1}$, R. Benzi$^{1}$, M. Bernaschi$^{2}$ S. Succi$^{2}$ \\
$^{1}$ Department of Physics and  INFN, University of ``Tor Vergata'', Via della Ricerca Scientifica 1, 00133 Rome, Italy\\
$^{2}$ Istituto per le Applicazioni del Calcolo CNR, Viale del Policlinico 137, 00161 Roma, Italy}

\begin{abstract} 
A systematic study for a single-specie lattice Boltzmann model with frustrated-short range  attractive and mid/long-range repulsive-interactions is presented.  The equilibrium analysis is performed along the guidelines proposed by [X. Shan, {\it Phys. Rev. E} {\bf 77}, 066702 (2008)]  and allows us to determine the surface tension density and the resulting disjoining pressure developing in a thin film when two interfaces overlap. Numerical simulations of confined flows are then performed with a multicomponent model and are successfully tested against the recent suggestion by Bocquet and coworkers on the existence of a cooperative  length underlying the non-local rheology of highly confined soft-glassy materials [Goyon et al., {\it Nature} {\bf 454}, 84­87 (2008); {\it Soft Matter} {\bf 6}, 2668-2678 (2010)].
\end{abstract}

\maketitle

\section{Introduction}
The dynamics of thin films and its impact on the rheology of surface-dominated flows has attracted a great deal of attention since long, due to its relevance from the theoretical point of view and its countless applications in science/engineering.  In particular, colloidal systems as foams (gas-liquid dispersions) and emulsions  (liquid-liquid dispersions) provide outstanding examples in point \cite{Larson,Coussot}. 

These systems are composed of one phase dispersed within another, and their overall rheology depends on the stability of individual films of intervening 'continuous' phases between two bubbles or drops of the dispersed phase. Thin liquid films can be classified as small thermodynamic (nanometric) systems. The special behavior of such systems is due to the absence of a bulk liquid core within the film: here different surface forces enter a competition and their outcome, depending on the specific physico-chemical conditions, determines a great variety of properties of the film. The disjoining pressure \cite{Derjaguin,Verwey,DerjaguinChuraev,Scheludko,Toshev} is surely one of the main thermodynamic characteristics of such systems. In foams and emulsions, the pressure in the dispersed phase is higher than the pressure  in the surrounding liquid, so that capillary suction occurs. The pressure difference is related to the radius of  the dispersed bubbles or droplets, according  to the Laplace equation.  The drainage of the intermediate thin films induced by the capillary pressure is slowed down and eventually inhibited whenever interactions between the film surfaces come into play. The disjoining pressure is due to these interaction forces between the two interfaces of the thin liquid film, as very nicely reviewed in \cite{Bergeron}.  This pressure acts perpendicular to the interfaces, thus balancing the capillary pressure, leading to a quasi-static equilibrium. Every interface represents a thin interfacial transition region whose intensive  thermodynamic properties deviate from those of the two neighbouring bulk phases. These transition regions develop as a natural consequence of the changes in the underlying molecular configurations, as one moves across a phase boundary. These molecular interactions give rise to macroscopic forces whenever any two phases approach each other, while an intervening third phase, separating them, gets thinner. 
 
The above situation portrays a highly complex scenario, which sets a challenge to most advanced numerical methods.  As typical of soft-glassy materials, this configures a no-man's land in which a continuum description becomes questionable on fundamental grounds, while a molecular approach still falls short of providing sufficient power to reach spatial and especially temporal scales of experimental interest. This no-man's land offers a perfect hunting ground for mesoscopic methods, working at the interface between continuum and molecular dynamics.

In the following we provide a detailed account of one such method, the Lattice Boltzmann technique for non-ideal fluids.

\section{Lattice Kinetic Theory for non-ideal fluids}

Hereafter, we shall consider an isothermal model at a fixed temperature $c_s^2=k_BT/m$ described by  the dynamics of a mesoscopic lattice system.  The motion of the fluid is described by a set of discrete single-particle distribution functions $f_l$: $l = 1 , . . . , d$, obeying the following dimensionless, velocity-discretized Boltzmann equation 
\be\label{LB}
f_l({\bm x}+{\bm c}_l,t+1)-f_l({\bm x},t)=-\frac{1}{\tau}\left(f_l({\bm x},t)-f^{(eq)}_l({\rho},{\bm u})\right)
\ee
where ${\bm x}$ and $t$ are spatial coordinates and time and $\{ {\bm c}_l:l=1,...,d \}$ the set of discrete velocities that coincide with the abscissas of a Gauss-Hermite quadrature in velocity space \cite{Shan06,Grad49}.  The quantity $f_l({\bm x},t)$ is essentially the countinuous single-particle distribution function $f({\bm x},{\bm v},t)$ evaluated for the velocity ${\bm v}={\bm c}_l$.   The right hand side describes the collisional relaxation of the probability distribution function  towards a local equilibrium distribution $f_l^{(eq)}({\rho},{\bm u})$.  By definition, the representative mesoscale particle collects a large number  molecules, i.e. all molecules contained in a unit cell of the lattice. Integration in momentum space provides the macroscopic fluid quantities, such as density and momentum
$$
\rho=\sum_{l=0}^{d} f_l \hspace{.2in} \rho {\bm u}=\sum_{l=0}^{d} f_l {\bm c}_l.
$$
Large scale momentum and energy conservation are secured, once the collisional 
kernel is designed for zero projection on the corresponding kinetic moments.  
In particular, when the system is close to equilibrium, the second-order 
tensor $\sum_{l=0}^{d} f_l {\bm c}_{l}^i {\bm c}_{l}^{j}$ reveals the momentum flux for large-scale hydrodynamics \cite{Benzi92,Chen98,Gladrow00}. 
For an ideal gas with local Maxwellian equilibrium, one finds
$$
\sum_{l=0}^{d} {\bm c}_{l}^i {\bm c}_{l}^{j}f_l \approx \sum_{l=0}^{d} {\bm c}_{l}^i {\bm c}_{l}^{j}f_l^{(eq)}=c_s^2 \rho \delta_{ij}+\rho {\bm u}_{i}{\bm u}_{j}
$$
where the first term on the right hand side represents the pressure of the ideal gas. In the presence of molecular interactions, such pressure receives non-ideal contributions. It is known that many intermolecular potentials can be taken in Lennard-Jones forms \cite{Rowlinson,Koplik}, namely a short-range strongly repulsive core (excluded-volume effect)  as combined with a long-range weakly attractive tail. However, when particles are located at fixed lattice points, and move with a uniform time-step, the modelling of the short-range molecular interactions has to be handled with care.  In the early papers by Shan \& Chen \cite{SC93,SC94}, the role of the excluded volume has been embedded directly into some effective density (or pseudopotential) $\psi(\rho)$. This pseudopotential may also be viewed as a generalized density, obeying the general properties of converging to the physical density in the low-density limit $\rho \rightarrow 0$, and saturating to a constant value at large densities. When interactions are proportional to inhomogeneities of $\psi$, the saturation naturally prevents mass collapse when density is going above a prescribed threshold.  More specifically, the force experienced by particles at ${\bm x}$ as due with interaction with particles at ${\bm z}$, takes the following form:
$$
{\bm F}_i({\bm x},{\bm z})={\cal G}_i(|{\bm x}-{\bm z}|)\psi({\bm x})\psi({\bm z}).
$$
where the subscript $i$ labels the spatial coordinate. For fast-decaying forces, when the sites interacting with the particles on ${\bm x}$ are limited to $N$ neighbors, not necessarily the nearest ones, the total force exerted on particles at ${\bm x}$ is obtained by summing over all ${\bm z}$. Therefore, given a limited set of links ${\bm c}_l$ \footnote{In principle not necessarily the same as those involved in the lattice Boltzmann dynamics} and requiring that the interaction be isotropic (i.e. that $|{\bm x}-{\bm z}|=|{\bm c}_l|$ carries the same interaction 
strength) we write
\be\label{FORCEa}
{\bm F}_i=-{\cal G}\psi({\bm x})\sum_{l=1}^{N} W(|{\bm c}_l|^2)\psi({\bm x}+{\bm c}_l){\bm c}^i_l
\ee
where ${\cal G}$ is a constant of proportionality dictating the overall strength of the non ideal interactions (${\cal G}<0$ encoding attractive interactions). 
Due to isotropy, the weights $W(|{\bm c}_l|^2)$ depend only on the square magnitude of the link. We wish to point out that the discrete velocities ${\bm c}_l$ can be identified with the discrete links of the lattice since the time-step is taken as a unit value throughout.

In the lattice Boltzmann schemes, the force is usually implemented via a shift \cite{SC93,SC94} of the  velocity field in the equilibrium distribution function (drifting Maxwellian)
$$
{\bm u}^{(eq)}_i \rightarrow {\bm u}^{(eq)}_i+\tau \frac{{\bm F}_i}{\rho}.
$$
Taylor expansion of the forcing field in (\ref{FORCEa}) delivers 
\be
{\bm F}_i \approx -{\cal G}\psi({\bm x})\left( ({\bm \nabla}^{(1)} \psi \cdot {\bm E}^{(2)})_i + ({\bm \nabla}^{(3)} \psi \cdot {\bm E}^{(4)})_i + ...\right)
\ee
where 
\be\label{tensorial}
{\bm E}^{(n)}={\bm E}^{(n)}_{i_1,i_2,...,i_n}=\sum_{l=1}^{N} W(|{\bm c}_l|^2){\bm c}_l^{i_1}{\bm c}_l^{i_2}...{\bm c}_l^{i_n}
\ee
is the generic $n$-th order tensor. With a given set of lattice vectors, it is therefore highly desirable to obtain  the finite-difference gradient operator with the highest possible degree of isotropy. This reduces to the problem of deriving the weights yielding the highest isotropic ${\bm E}^{(n)}$. References \cite{Shan06,Sbragaglia07} give solutions for isotropy tensors up to ${\bm E}^{(10)}$ in both $2d$ and $3d$ cases.

\subsection{Competing interactions}\label{sec:competing}

Nearest-neighbor interactions in (\ref{FORCEa}) have been widely used to describe a  rich variety of complex flows \cite{SC93,SC94,Sbragaglia06, Hyvaluomaetal}.  Given the pseudopotential $\psi(\rho)$, a suitable choice of ${\cal G}<0$ permits to  describe phase transitions and stable liquid-gas interfaces.  Here we detail the interface properties of a model with {\it frustrated} nearest-neighbor and next-to-nearest neighbor interactions.  The details reported hereafter refer to a two dimensional $(x,z)$ model and extensions to  three dimensional cases can be developed along the lines of references \cite{Sbragaglia07,Shan06}.  The 'short' range interactions encode interparticle attraction  (with strength coefficient ${\cal G}_1<0$ and weights $w(|{\bm c}_l|^2)$)  and extend up to velocities with $|{\bm c}_l|^2=4$; a competing repulsive 'long' range interaction  (with strength ${\cal G}_2>0$ and weights $p(|{\bm c}_l|^2)$) extends up to velocities with $|{\bm c}_l|^2=8$.  
In equations:
\be\label{FORCE}
\begin{split}
{\bm F}_i= -& {\cal G}_1  \psi({\bm x})\sum_{l=1-12} w(|{\bm c}_l|^2)\psi({\bm x}+{\bm c}_l){\bm c}^i_l\\
-& {\cal G}_2\psi({\bm x})\sum_{l=1-24} p(|{\bm c}_l|^2) \psi({\bm x}+{\bm c}_l){\bm c}^i_l.
\end{split}
\ee
The choice to extend the attractive interactions up to $|{\bm c}_l|^2=8$ instead of  $|{\bm c}_l|^2=4$ (as considered in \cite{CHEM09}) responds to the intent of preserving the isotropy of the sixth order tensors for both interactions.  

\begin{figure}
\centering
\includegraphics[scale=0.3]{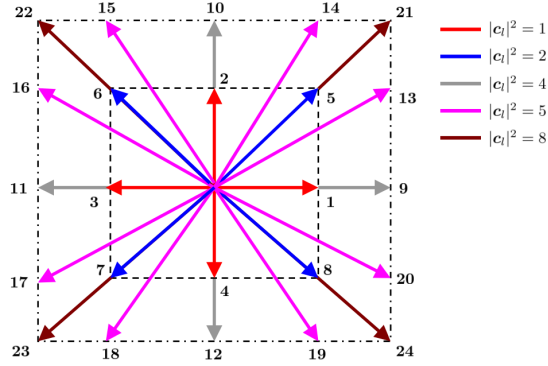}
\caption{The discrete $25$-speeds lattice. Both belts are shown, along with the corresponding discrete velocities. The usual nearest-neighbor Shan-Chen \cite{SC93,SC94} model corresponds to the first $8$ velocities (first belt). The competing frustrated-short range attractive and mid/long-range repulsive-interactions are obtained with two-belts interactions. Both interactions have been chosen in such a way as to preserve isotropy of the $6$-th order tensors in the velocity fields. \label{fig:1}}
\end{figure}

\section{The Stress Field on The lattice}\label{sec:stress}

Once the mechanical model for the lattice interactions is laid down, see equation (\ref{FORCE}), it becomes crucial to determine the associated {\it stress} or {\it pressure tensor} \cite{Sbragaglia07,Shan08}, responsible for  mechanical balance at the interface. Remarkably, an exact lattice theory is available \cite{Shan08} and in  this section we briefly recall its essential features. The exact stress is given by
$$
{\bm \sigma}_{ij}=c_s^2 \rho \, {\bm \delta}_{ij}+{\bm \sigma}^{(int)}_{ij}+\frac{1}{\rho}\left(\tau-\frac{1}{2}\right)^2 {\bm F}_i {\bm F}_j
$$
where ${\bm \delta}$ is the unit tensor and $P_{id}(\rho)=c_s^2 \rho$ is the ideal pressure contribution. The interaction forces ${\bm F}$ are connected to the interaction stress ${\bm \sigma}^{(int)}_{ij}$ by  requiring that the force comes as the divergence of the stress tensor. Mathematically, this implies that the relation
\be\label{MECHANICAL}
\begin{split}
{\bm F}_i=& -{\cal G}_1\psi({\bm x})\sum_{l=1-12} w(|{\bm c}_l|^2) \psi({\bm x}+{\bm c}_l){\bm c}^i_l\\
& -{\cal G}_2\psi({\bm x})\sum_{l=1-24} p(|{\bm c}_l|^2) \psi({\bm x}+{\bm c}_l){\bm c}^i_l=-{\bm \nabla}_j  {\bm \sigma}^{(int)}_{ij}
\end{split}
\ee
must hold {\it exactly} on the lattice. To this aim, one considers the various directional force vectors  $\psi({\bm x}) \psi({\bm x}+{\bm c}_l){\bm c}_l$, and computes  their flux over the unit area.  The stress is then obtained by summing over all interacting links \cite{Shan08}. For the 'short' range interactions ($\ell=1-12$), one obtains:
\be\label{PT}
{\bm \sigma}^{(int)}_{ij}={\bm \sigma}^{(1)}_{ij}+{\bm \sigma}^{(2)}_{ij}+{\bm \sigma}^{(4)}_{ij}
\ee
where ${\bm \sigma}^{(\alpha)}_{ij}$ indicates the contribution of the shell with energy $|{\bm c}_l|^2=\alpha$ (see Appendix A for the explicit details). When next to nearest neighbor are included, the analytical details become slightly more involved, but  expressions similar to (\ref{PT}) are still obtained \cite{Shan08}.  In particular, for the two-belts interactions depicted in figure (\ref{fig:1}), the corresponding interaction stress reads as follows:
\be
\begin{split}
{\bm \sigma}^{(int)}_{ij}={\bm \sigma^{(1)}_{ij}}+{\bm \sigma^{(2)}_{ij}}+{\bm \sigma^{(4)}_{ij}}+{\bm \sigma^{(5)}_{ij}}+{\bm \sigma^{(8)}_{ij}}
\end{split}
\ee
where ${\bm \sigma}^{(1)}_{ij}$, ${\bm \sigma}^{(2)}_{ij}$ and ${\bm \sigma}^{(4)}_{ij}$ have been given before (we simply have to replace $w (|{\bm c}_l|^2)$ with $p (|{\bm c}_l|^2)$ and ${\cal G}_1$ with ${\cal G}_2$) while the extra terms ${\bm \sigma^{(5)}_{ij}}$, ${\bm \sigma^{(8)}_{ij}}$ are, again, detailed in appendix A. The 'bulk' contribution of the stress identifies the interaction contribution to the bulk equation of state which reads
\be\label{BULK}
P_b(\rho)=c_s^2 \rho+c_s^2 \frac{({\cal G}_1+{\cal G}_2)}{2} \psi^2(\rho).
\ee
Going to higher orders, we find contributions to surface forces, i.e. forces active at the interfaces separating the bulk phases. As usual in these situations, it is expedient to analyze a one dimensional problem: for a planar one dimensional interface extending from $x=-\infty$ to $x=+\infty$, the mismatch between the normal ($\sigma_{xx}$) and tangential ($\sigma_{zz}$) components of the interaction stress reads as follows:
\begin{equation}\label{DIS1}
\begin{split}
p_s(x)=&(\sigma_{xx}-\sigma_{zz})(x)=\\
&(\sigma^{(int)}_{xx}-\sigma^{(int)}_{zz})(x)+\frac{1}{\rho}\left(\tau-\frac{1}{2}\right)^2 {F}^2_x(x)
\end{split}
\end{equation}
where the term $(\sigma^{(int)}_{xx}-\sigma^{(int)}_{zz})(x)$ is exactly written on the lattice. By performing a Taylor expansion of the fields (see Appendix A), we get the various contributions to the total surface tension
\be
\begin{split}
\gamma& = \int_{-\infty}^{+\infty} p_s(x) dx=\int_{-\infty}^{+\infty}(\sigma^{(int)}_{xx}-\sigma^{(int)}_{zz})(x) dx= \nonumber \\
& \int_{-\infty}^{+\infty} \left(C_2 \left(\frac{d\psi}{dx} \right)^2+{C}_4 \left(\frac{d^2 \psi}{d x^2} \right)^2+\left(\tau-\frac{1}{2}\right)^2\frac{F^2_x(x)}{\rho(x)} \right) dx.
\end{split}
\ee
and
\be\label{C2}
{C}_2=-\frac{{\cal G}_1}{2} e_4(w)-\frac{{\cal G}_2}{2} e_4(p)
\ee
\be\label{C4}
{C}_4=\frac{{\cal G}_1}{4} e_6(w)+\frac{{\cal G}_2}{4} e_6(p).
\ee
The coefficients $e_4$, $e_6$ appearing in front of the coupling constants ${\cal G}_{1,2}$, may be associated with suitable momenta resulting from the definition (\ref{tensorial}). We remark that the coefficients in ${C}_2$ are also in agreement with the analysis presented by Shan \cite{Shan08}, regarding the surface tension effects coming from the expression of the exact stress (see also appendix B). In particular, by using the condition of isotropy of fourth-order tensors, one obtains 
$$
e_4(W)=W(1)+16 W(4) +18 W(5) \hspace{.2in} W=w,p
$$ 
$$
e_6(W)=\frac{W(1)}{6}+\frac{32}{3} W(4)+ 15 W(5)\hspace{.2in} W=w,p.
$$

\begin{figure}[t]
\includegraphics[scale=0.7]{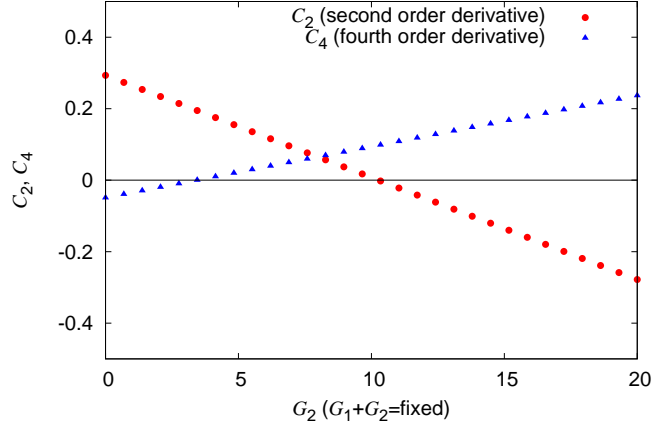}
\caption{The coefficients ${C}_2$, ${C}_4$ defined in equations (\ref{C2}) and (\ref{C4}). The coefficients  ${C}_2$ and ${C}_4$ control the surface stresses proportional to squared pseudopotential gradients and the higher order terms (squared pseudopotential second derivative). A negative  ${C}_2$ at positive ${C}_4$ marks the onset of the emergence for a positive disjoining pressure. \label{fig:2}}
\end{figure}

The fourth and sixth order tensors are positive defined $e_4(W),e_6(W) >0$ and, 
by properly choosing ${\cal G}_1<0$ and ${\cal G}_2>0$ (see figure \ref{fig:2}), one can realize the condition ${C}_2<0$ and ${C}_4>0$, i.e. the squared gradient terms contribute negative terms  so that a very steep interface would be favored.  The higher order terms counterbalance this with a positive (${C}_4>0$) bending rigidity.  In fact, by assuming (mechanical equilibrium) that $c_s^2 \frac{d \rho(x)}{dx}=F_x(x)$, we find
\be\label{PS}
\begin{split}
\gamma = &\int_{-\infty}^{+\infty} p_s(x) dx =\int_{-\infty}^{+\infty}(\sigma^{(int)}_{xx}-\sigma^{(int)}_{zz})(x) dx= \nonumber \\
&\int_{-\infty}^{+\infty} \left( {C}_2+\frac{c_s^4}{\rho (\psi^{\prime}(\rho))^2}\left(\tau-\frac{1}{2}\right)^2 \right) \left(\frac{d\psi}{dx} \right)^2 dx+\\
&\int_{-\infty}^{+\infty}{C}_4 \left(\frac{d^2 \psi}{d x^2} \right)^2   dx
\end{split}
\ee
This expression allows us to make a direct link with interfacial models of micro emulsions  \cite{Helfrich,Safran}, which assume the presence of an implicit surfactant monolayer, whose elastic energy is given by a suitable Helfrich Bending-Hamiltonian \cite{Helfrich}. In fact, the surfactant degrees of freedom (say, the surfactant concentration) are integrated out and do not appear explicitly in the free-energy forms (\ref{PS}). Their presence manifests through the specific form of the coefficients  upfronting the various gradient terms.  Following Gompper and Zschocke \cite{Gompper}, and drawing an intriguing parallel between  the pseudopotential gradients in our equation (\ref{PS}) and the order parameter  gradients reported in equations (16a)-(16e) of \cite{Gompper}, it is possible to extract the precise expression for spontaneous curvature radius, the bending rigidity  and the saddle-splay modulus, whose details shall be given in a future publication.   

A few comments on the crucial role of the finite $\tau$ contributions in (\ref{PS}) are in order.  On the assumption that the coefficient in front of $(\frac{d\Psi}{dx})^2$ in the equation (\ref{PS}) be a constant, one can express the surface tension as a simple integral in Fourier-space as $\gamma \propto \int \psi(k) G(k)  \psi(-k) dk$, where the kernel is given by $G(k)={C}_2 k^2 + {C}_4 k^4$.  Whenever ${C}_2<0$ and ${C}_4>0$, the kernel $G(k)$ shows a minimum at a  finite wavenumber $k_0$, withnessing an instability at the interface. This instability generates a pattern domain, with a characteristic  wavenumber $k_0 \approx \sqrt{-{C}_2/ 2{C}_4} $ \cite{Seul95}.  


Our analysis, and specifically the expression (\ref{PS}), illuminates the basic reason of such failure.  The point is that the term upfronting $(\frac{d\Psi}{dx})^2$ is generally a function of space and, in fact, a {\it strongly} varying function as the two interfaces come together.  Under such conditions, the surface tension can no longer be encoded within a local kernel $G(k)$, but requires a full convolution in $k$-space instead. More specifically, the discrete forcing correction proportional to $(\tau-1/2)$  segregates the instability within the interface and prevents it from developing  outside the layer.  The square of the pseudopotential gradient leads to a spatial modulation  along the interface, with regions characterized by both negative and positive signs, the overall surface tension being left small but positive.  This 'localized' instability triggers a density kink in the proximity of the  bulk phases, which is directly responsible for the emergence of a positive disjoining pressure when two interfaces tend to overlap. In the next sections we will first detail the emergence of the positive disjoining pressure when a suitable degree of frustration from the competing interactions is chosen. Then, we will show how such positive disjoining pressure is  directly related to the non-linear rheological behavior of the fluid mixture. 

\subsection{The case of two near-interfaces: emergence of the disjoining pressure}

\begin{figure}
\centering
\includegraphics[scale=0.65]{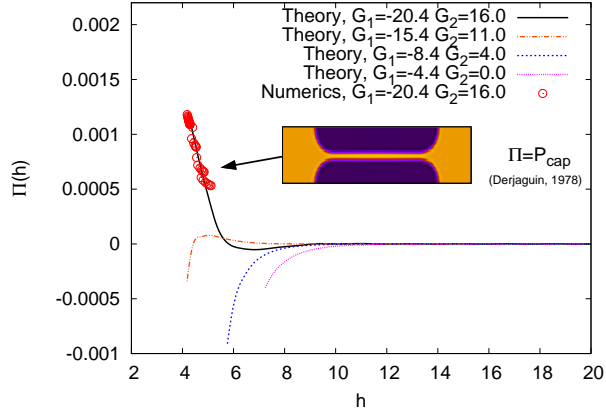}
\caption{A direct measurement of the disjoining pressure. Two hemispherical bubbles are faced one against each other and form a thin liquid film between them.  The film is stabilized against coalescence due to the choice of the parameters (${\cal G}_1=-20.4$, ${\cal G}_2=16.0$) corresponding to a positive disjoining pressure.  The mechanical definition of Derjaguin \& Churaev \cite{DerjaguinChuraev} is  then applied and the disjoining pressure is measured from  the capillary pressure at the curved interface. Different theoretical predictions are shown for different values of the coupling parameters at fixed bulk pressure (\ref{BULK}). \label{fig:3}} 
\end{figure}

The disjoining pressure $\Pi$ is a very basic thermodynamic quantity of thin liquid films.  Nevertheless, a rigorous definition of this quantity has remained elusive for a long time. No bulk liquid core exists within the thin film, and this inhomogeneity implies that the mechanical state of such film should be defined not in the terms  of a scalar pressure, but rather in terms of a pressure tensor, with  separate normal and tangential components.  Given the results presented in the previous section, we are in the position to control exactly the emergence of the disjoining pressure.  To this end, we consider {\it two} non-ideal interfaces, separated by the distance  $h$. Following Bergeron \cite{Bergeron}, the overall film tension reads as follows:
\be\label{doppiasigma}
\gamma_f=2 \gamma(h=\infty)-\int_{\infty}^{h} \Pi \, dh +\Pi h=2 \gamma(h=\infty)+\int_{\Pi(h=\infty)}^{\Pi(h)} h \, d \Pi
\ee
where $\gamma(h=\infty)$ is the bulk value of the surface tension and $\gamma_f$ is the overall film tension, whose expression is known in terms of the mismatch between the normal and tangential components of the pressure tensor \cite{Toshev,Derjaguin}. Based on the analysis developed in the previous section and the help of equation (\ref{doppiasigma}), we determine (see figure \ref{fig:3}) the disjoining pressure for different values of the coupling parameters at fixed bulk pressure (\ref{BULK}), i.e. at fixed ${\cal G}_1+{\cal G}_2$. To test the theoretical prediction, we performed the following numerical simulations: two hemispherical bubbles are faced one against each other, so as to form a thin liquid film inbetween. The film is stabilized against coalescence, due to the choice of the parameters (${\cal G}_1=-20.4$, ${\cal G}_2=16.0$) corresponding to a positive disjoining pressure. Full periodic boundary conditions are then applied. Following Derjaguin \& Churaev \cite{DerjaguinChuraev}, we note that in mechanical equilibrium the disjoining pressure must be equal to the difference existing between the component of the pressure tensor in the interlayer (a constant at machine precision in our case) and the pressure set up in the bulk of the phase from which it has been formed by thinning out. This means that the capillary pressure between the bulk phases in the 'bubble' and outer regions must equal the disjoining pressure. By varying the the radii of the bubbles, we are able to capture different disjoining pressures for various widths $h$. The results are displayed in figure \ref{fig:3} and reveal satisfactory agreement with the theoretical prediction.

\section{Numerical simulations of confined pressure-driven flows at different packing-fractions}

The above lattice kinetic theory offers a very powerful and efficient computational tool to investigate a variety of complex dynamic phenomena occurring in soft-glassy flows, such as anomalous relaxation, dynamic arrest and non-linear rheology in general \cite{CHEM09,EPL10}.\\ 
Recently, Bocquet and collaborators \cite{Goyon08,Goyon10} proposed a theoretical framework able to control finite-size effects in the rheological behaviour of confined systems. Based on the idea that flow occurs via a succession of reversible elastic deformations and local irreversible plastic rearrangements associated with a microscopic yield stress, they developed a theory accounting for a non-local, long-ranged, elastic relaxation of the stress over the system. Evidence has been provided that this long-range relaxation might be traced to the onset of a so-called {\it cooperative length} $\zeta$.  More precisely, the cooperative length fixes the spatial distribution of the  fluidity $f=\dot{\gamma}/\sigma$ (basically the inverse effective viscosity, with $\sigma$ the stress and $\dot{\gamma}$ the strain rate)  through the following diffusion equation: 
\be\label{fluidity}
\zeta \Delta f = (f-f_{b})
\ee
with $f_b$ the 'bulk' fluidity \cite{Goyon08} so that, whenever the size of the domain becomes comparable to $\zeta$, non-local effects can no longer be neglected. The authors go on, by giving the expression of the cooperative length as a function of the packing fraction $\Phi=V_d/V$, where $V_d$ is the volume of the dispersed  phase over the total volume available, $V$.  Clearly, in the limit $\Phi \rightarrow 1$, the average intergap distance between the  dispersed droplets tends to zero, and thin-film effects start dominating the picture. 

\begin{figure*}\begin{center}
\includegraphics[scale=0.34]{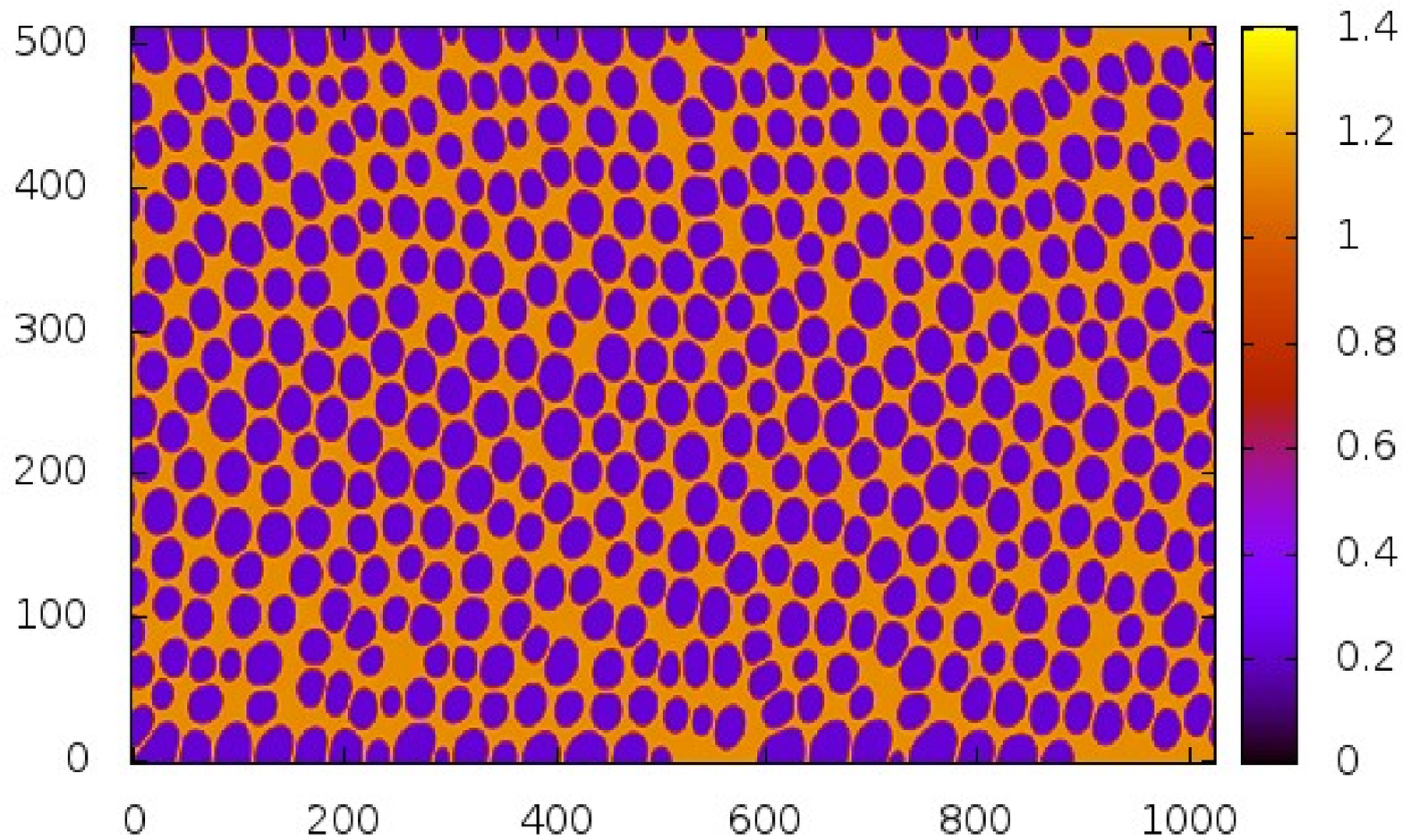}
\includegraphics[scale=0.34]{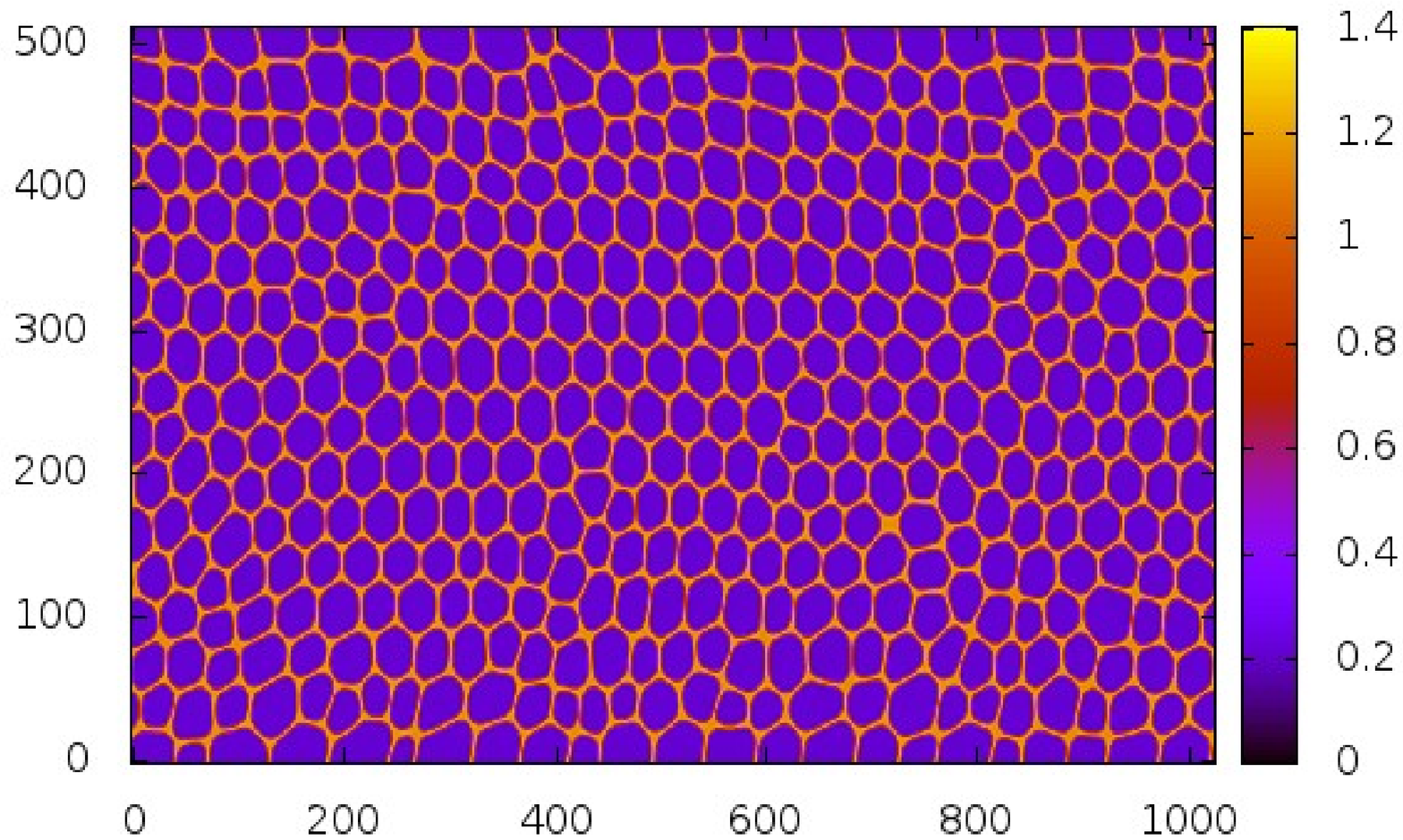}
\caption{Two typical bi-liquid emulsion like configurations of the binary fluid for two different mass packing fractions ($\Phi_{mass}=0.536$ and $\Phi_{mass}=0.628$). Blue and yellow code for high and low density of fluid $A$, respectively. Hence, the blue spots in the left panel represent dispersed liquid droplets. Varying the packing fraction we can simulate a 'foamlike' network of thin film layers within dispersed droplets (right panel). The different configurations are obtained by changing  the overall mass ratio between fluid $A$ and $B$.\label{fig:4}}
\end{center}\end{figure*}
  
It is therefore of great interest to investigate whether such thin-film effects are indeed captured by our lattice kinetic model and, more precisely, whether such effects can be interpreted in terms of the notion of cooperative length. To this purpose, we consider a binary mixture of fluids $A$ and $B$ (see figure \ref{fig:4}), each described  by a discrete kinetic Boltzmann distribution function $f_{l,s}({\bm x},t)$, yielding the probability of finding a representative particle of fluid $s=A,B$ at position ${\bm x}$ and time $t$. The distribution functions of the two fluids evolve under the effect of free-streaming and local two-body collisions as seen in (\ref{LB}). As for the interparticle forces, first of all, a repulsive force (whose strength is proportional to ${\cal G}_{AB}$) between the two fluids ensures phase separation \cite{CHEM09}. Furthermore, both fluids are also subject to short range attraction  (with strength parameters ${\cal G}_{AA,1}$, ${\cal G}_{BB,1}$), and long-range (with strength parameters ${\cal G}_{AA,2}$, ${\cal G}_{BB,2}$) intra-species ($AA$, $BB$) repulsion, i.e. the very same mechanism detailed in section \ref{sec:stress}.\\
The simulations pertain to a planar Poiseuille flow on a $L \times H = 1040 \times 1040$ computational domain. We prepare the system with a collection of polydisperse ``packings'' of fluid $A$ separated by layers of fluid $B$ and being stabilized by a positive disjoining pressure (see also figure \ref{fig:4}). This resembles a bi-liquid emulsion system. In both cases \footnote{All simulation results given in lattice Boltzmann units.}, the $A$ rich ($B$ rich) region has approximately $\rho_A=1.2$, $\rho_B=0.2$ ($\rho_A=0.2$, $\rho_B=1.2$). The hydrodynamic viscous ratio between $A$ rich and $B$ rich regions is $1$. The coupling parameters are ${\cal G}_{AA,1}=-9.0$, ${\cal G}_{AA,2}=8.1$ ${\cal G}_{BB,1}=-8.0$, ${\cal G}_{BB,2}=7.1$, ${\cal G}_{AB}=0.587$. The form of the pseudopotential used is $\psi_{A,B}(\rho)=1-e^{-\rho_{A,B}/\rho_0}$ with  $\rho_0=0.83$ \cite{CHEM09}.\\ We also applied a volume force with a constant pressure gradient in the stream-flow $x$ direction $\Delta P/L$ and set a no-slip boundary condition for the velocity field at the boundaries located in $z=0$ and $z=H$. Each simulation, spanning multi-million time steps for every single set of parameters, takes about $30$ hours on a $2050$ Nvidia-GPU. The 2050 Nvidia-GPU features 448 cores grouped in 14 Streaming Multiprocessors running at 1.15 GHz. The code relies on the CUDA 4.0 and offers a speedup in excess of one order of magnitude with respect to a highly tuned (multi-core) CPU version \cite{ourPRE}.

\begin{figure*}[t]
\centering
\includegraphics[scale=0.6]{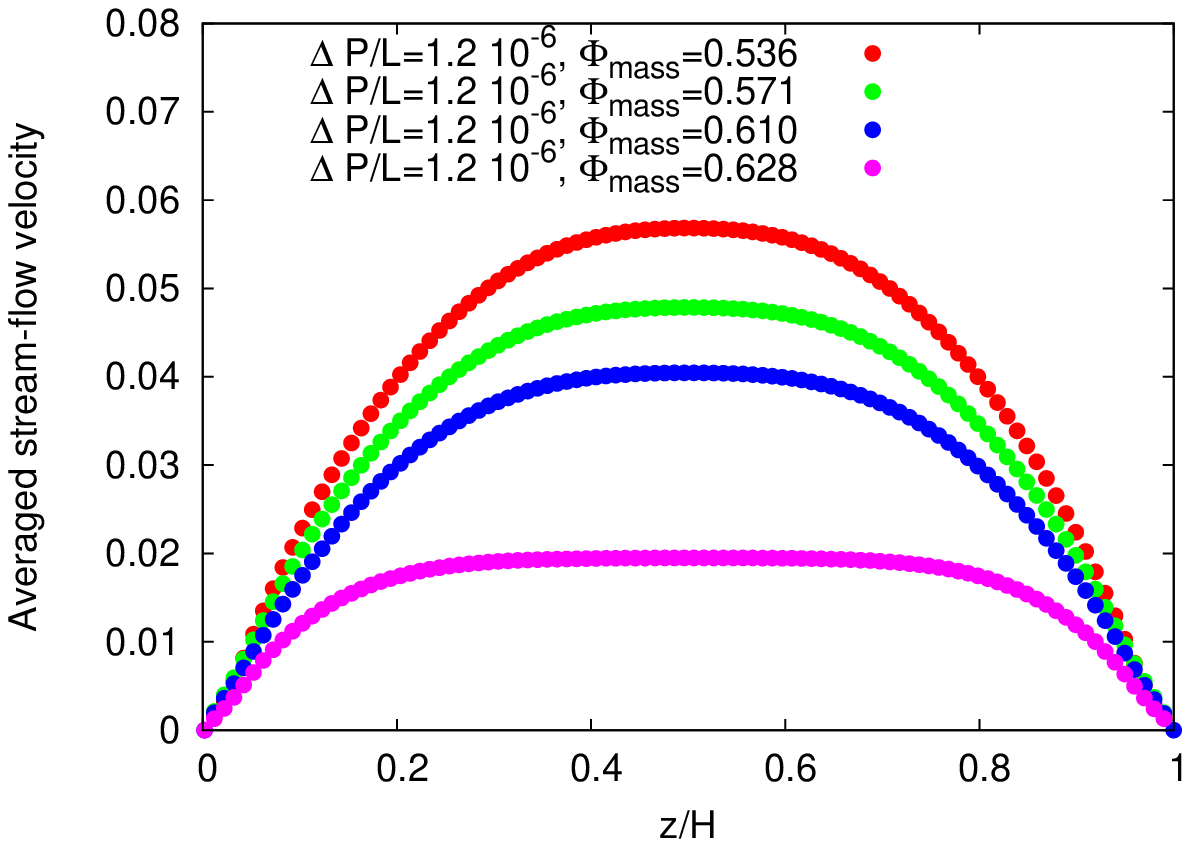}
\includegraphics[scale=0.6]{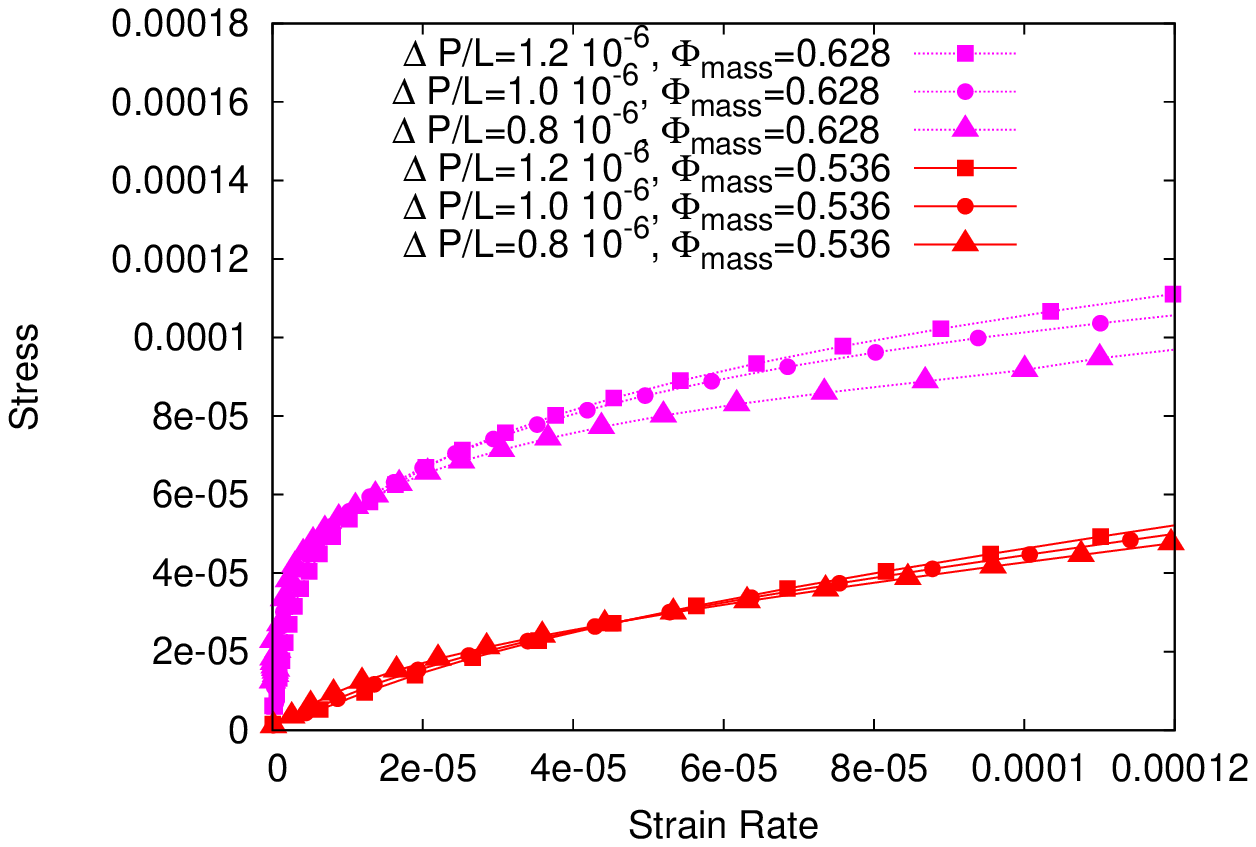}
\caption{Non-local rheological effects in pressure driven flows. Left panel: we report the time average of the stream-flow velocity profile $\langle \bar{u}_x(z) \rangle$ along the cross-flow coordinate $z/H$, at a given value of the pressure gradient $\Delta P/L$, for different values of the mass packing fraction $\Phi_{mass}$. The right panel shows the local rheological curves, i.e. the local stress {\it vs.} the local strain rate obtained from the time average of the stream-flow velocity profile for the mass packing fractions $\Phi_{mass}=0.537$, $\Phi_{mass}=0.628$ and different values of the pressure gradient $\Delta P/L$. The packing fraction is an indication of how much the colloidal droplets are packed.  \label{fig:5}}
\end{figure*}

In figure \ref{fig:5}, left panel, we report the time and stream-flow average of the stream-flow velocity $\langle {\bar{u}}_x(z) \rangle$ (where $\bar{...}$ denotes the average in the stream-flow direction and the brackets denote averaging over time) along the cross-flow coordinate $z/H$, at a given value of the pressure gradient $\Delta P/L$, for different values of the mass packing fraction $\Phi_{mass}$. The right panel shows the local rheological curves, i.e. the local stress {\it vs.} the local strain rate obtained from the averaged stream-flow velocity profile for the mass packing fractions $\Phi_{mass}=0.537$ and  $\Phi_{mass}=0.628$ and different pressure gradients. To compute the packing fraction, we use the mass instead of the volume, because the latter would be hard to measure exactly due to finite-width interface overlapping effects, while the former is strictly  dictated by the initial conditions, since the  mass of species $A$ and $B$ is conserved separately for both. Overall, figure \ref{fig:5} highlights the emergence of a non linear rheology with associated yield stress at increasing mass packing fraction. More precisely, when the mass packing fraction is increased, local rheological curves obtained for different pressure gradients (see right panel) are scattered and do not collapse on a single rheological curve, a fact has been rationalized in the theoretical framework developed in \cite{Goyon08,Goyon10}

\begin{figure*}[t]
\begin{center}
\includegraphics[scale=0.6]{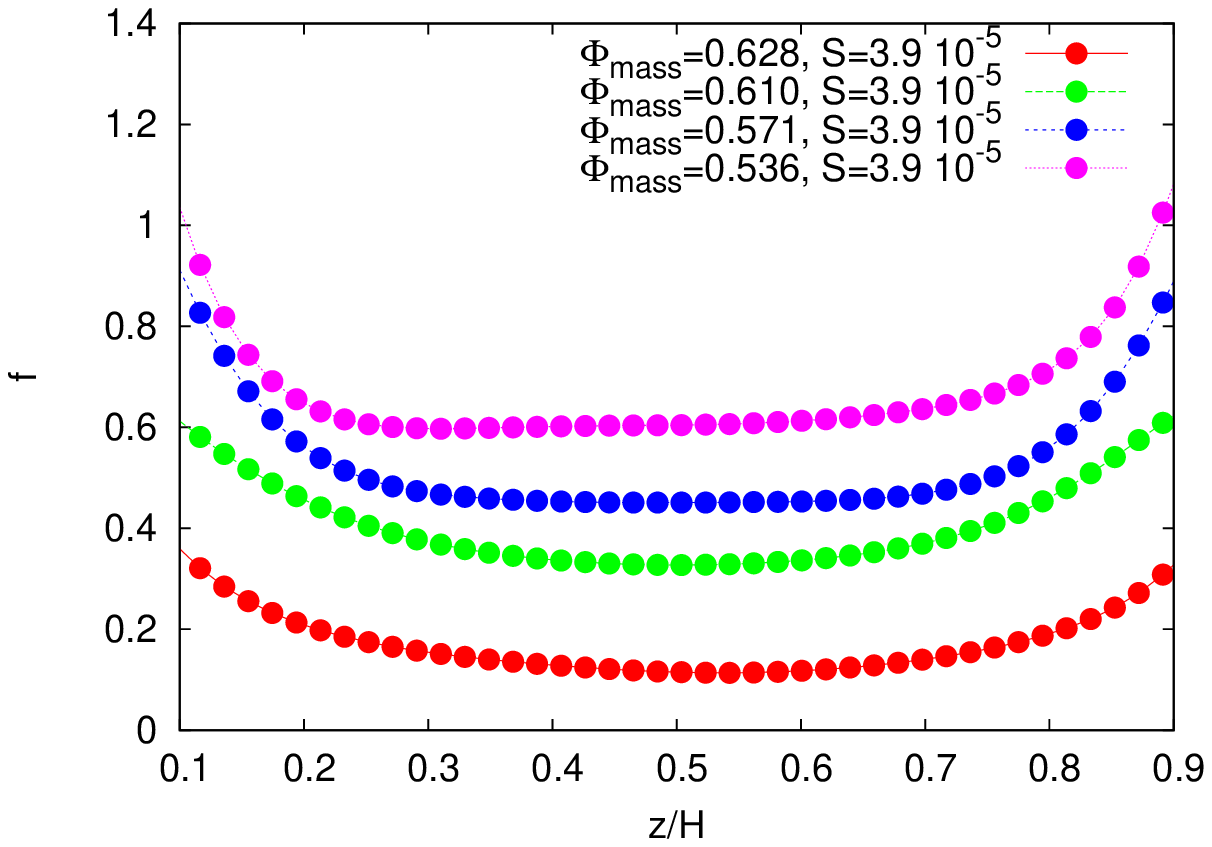}
\includegraphics[scale=0.6]{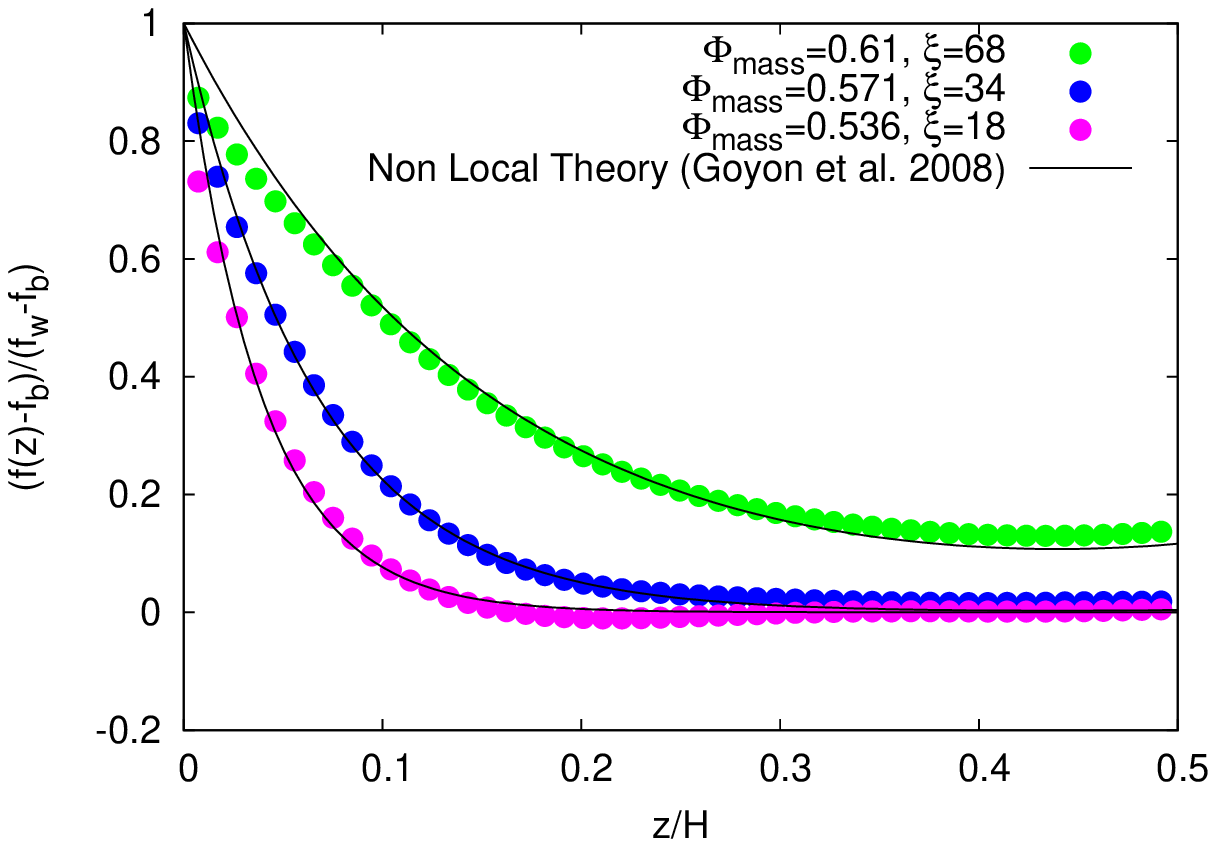}
\caption{The local fluidity field in a planar Couette cell where the material is sheared between two parallel walls (with velocity $\pm U_w$ ) separated by the distance $H$. Left panel: we show the local fluidity $f(z)$, as a function of the cross-flow coordinate at a fixed strain rate $S=2U_w/H$, for different values of the mass packing fraction. Right panel: we report the relative fluidity departure from the bulk value as a function of the distance from the wall located at $z=0$. The theoretical prediction given by equations (\ref{fluidity}) and (\ref{predict}) is also reported and reveals the different cooperative lengths $\zeta$ associated with different mass packing fractions. \label{fig:6}}
\end{center}
\end{figure*}

To make an even closer contact with the prediction of equation (\ref{fluidity}), we have run simulations in a planar Couette cell, shearing the material between two parallel walls separated by the distance $H$ by applying a symmetric velocity $\pm U_W$ to the walls. As explained in \cite{Goyon10}, this is a useful setup where one can test the prediction coming from equation (\ref{fluidity}), since the mean shear stress is spatially homogeneous, i.e. $\langle \bar{\sigma}(z) \rangle=\sigma_0$. In particular, the integration of equation (\ref{fluidity}) between the wall region ($w$) and a generic $z$, delivers the following result (see also equation (7) in \cite{Goyon10}):
$$
f(z)=\left[f_b(\sigma_0)+(f_w-f_b(\sigma_0)) \frac{\cosh((z-H/2)/\zeta)}{\cosh(H/2\zeta))} \right]
$$
or equivalently
\be\label{predict}
\frac{f(z)-f_b(\sigma_0)}{f_w-f_b(\sigma_0)}=\frac{\cosh((z-H/2)/\zeta)}{\cosh(H/2\zeta))}
\ee
where $f_w$ is the wall fluidity that we can measure directly in the numerical simulations. In the numerical simulations, we dump the stream-flow averaged velocity profile ${\bar{u}}_x(z)$ and the stress ${\bar{\sigma}}_x(z)$. The fluidity field is directly obtained from the ratio between the gradient of ${\bar{u}}_x(z)$ and the stress. Such value is then averaged in time.  In figure \ref{fig:6}, left panel, we show such fluidity $f(z)$, as a function of the cross-flow coordinate at a fixed strain rate $S=2U_w/H$, for different values of the mass packing fraction. The figure clearly reveals a sharp decrease of the local fluidity away from the wall, where the fluid flows like a liquid, towards the centerline, where the fluid flow is significantly inhibited. The right panel, which reports the relative fluidity departure from the bulk value as a function of the distance from the wall (located at $z=0$), makes this observation even sharper. In particular, the relative fluidity well adapts to the functional behaviour predicted by the cooperative-length model in equation (\ref{predict}), with an impressive {\it quantitative} agreement between the numerical and analytical data. The corresponding cooperative length $\zeta$ is increasing as the mass packing fraction increases: from $\zeta=18$ ($\Phi_{mass}=0.536$) to $\zeta=68$ ($\Phi_{mass}=0.610$). This figures provides a compelling evidence that the present kinetic model with competing interaction does indeed support the notion of a cooperative length in association with non-linear rheology of soft-glassy materials.    
  
\begin{figure*}
\begin{center}
\includegraphics[scale=0.6]{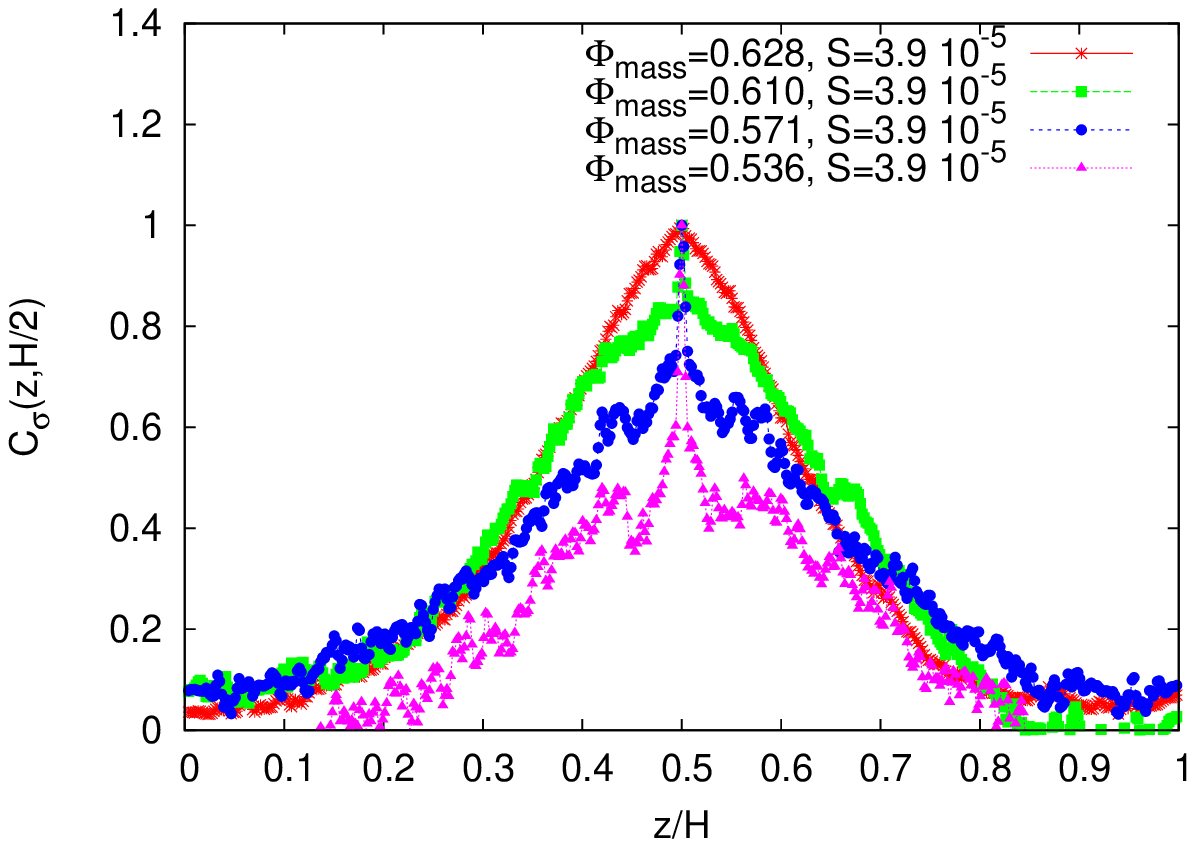}
\includegraphics[scale=0.6]{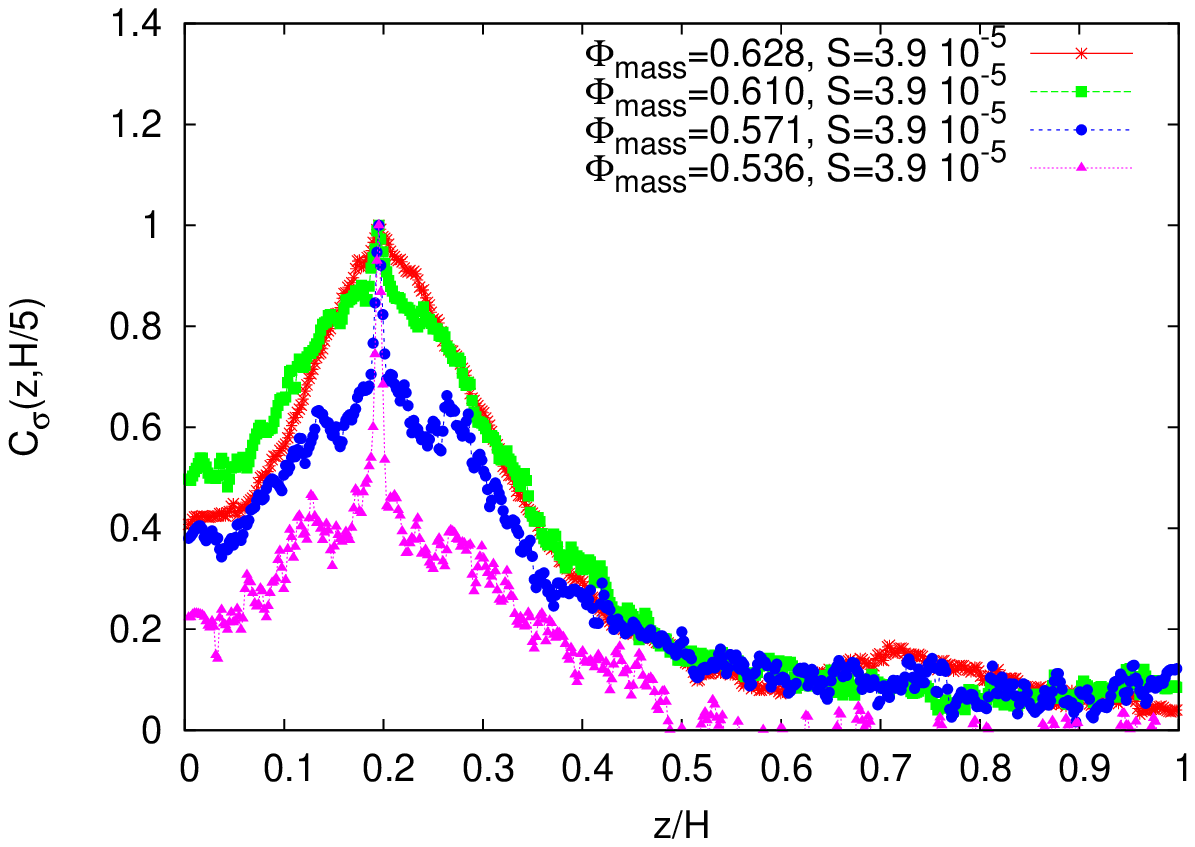}
\caption{The stress correlation $C_{\sigma}(z,z_0)$ as defined in equation (\ref{correlator}) at the mid-channel $z_0=H/2$ (left) and for $z_0=H/5$ (right). Different mass packing fractions $\Phi_{mass}$ are considered. The peak is a measure of Newtonian behavior and fades away at increasing mass packing fraction. \label{fig:7}}
\end{center}
\end{figure*}

It is also of interest to monitor the spatial behavior of the stress correlator
\be\label{correlator}
C_{\sigma}(z,z_0) =  \frac{ \langle \bar{\sigma} (z_0) \; \bar{\sigma} (z) \rangle-\langle \bar{\sigma} (z_0) \rangle \langle \bar{\sigma} (z) \rangle}{\langle \bar{\sigma}^2 (z_0) \rangle-\langle \bar{\sigma} (z_0) \rangle^2 }.
\ee
In figure \ref{fig:7} we report the stress correlation as a function of the cross-flow coordinate $z$ at a given value of the strain rate $S=2U_w/H$ and different values of the mass packing fraction $\Phi_{mass}$. The figure clearly shows that by increasing the mass packing fraction, the profile $C_{\sigma}(z,z_0)$ looses the Dirac's delta peak associated with a Newtonian behavior without spatial fluctuations (just a single value of $\sigma$ throughout the fluid). Instead, the Dirac peak is replaced by a continuous distribution of values. 

\section{Conclusions and outlook}

Summarizing, we have provided details of the way how supramolecular (dispersion)  forces emerge from a lattice kinetic model with frustrated interactions. The effects of these forces have been illustrated for the case of pressure-driven and sheared soft-glassy flows (emulsions) under confinement. The simulations provide clear evidence of non-Newtonian behavior  at increasing packing fractions, in close quantitative agreement with the cooperative-length model recently proposed in the literature \cite{Goyon08,Goyon10}. Further effects related to the onset of disjoining forces on the non-linear rheology of the soft-glassy flows under strong confinement will be reported in future publications. 

M. S. acknowledges support from DROEMU-FP7 IDEAS Contract No. 279004.

\section{Appendix A}
The relevant terms for the 'short' range interactions are:
\be
{\bm \sigma}^{(1)}_{ij}=\frac{{\cal G}_1}{2} \psi({\bm x}) \sum_{l=1-4} w(1) \psi({\bm x}+{\bm c}_l) {\bm c}^i_l {\bm c}^j_l
\ee
\be
{\bm \sigma}^{(2)}_{ij}=\frac{{\cal G}_1}{2} \psi({\bm x}) \sum_{l=5-8} w(2) \psi({\bm x}+{\bm c}_l) {\bm c}^i_l {\bm c}^j_l
\ee
\be
\begin{split}
{\bm \sigma}^{(4)}_{ij}=&\frac{{\cal G}_1}{4} \sum_{\ell=9-12} w(4)  \psi({\bm x}) \psi({\bm x}+{\bm c}_{\ell}){\bm c}_{\ell}^i {\bm c}_{\ell}^j\\
+&\frac{{\cal G}_1}{4} \sum_{\ell=9-12} w(4) \psi\left({\bm x}-\frac{{\bm c}_{\ell}}{2}\right)\psi\left({\bm x}+\frac{{\bm c}_{\ell}}{2}\right)   {\bm c}_{\ell}^i {\bm c}_{\ell}^j
\end{split}
\ee
In the case of 'long' range interactions,  ${\bm \sigma}^{(1)}_{ij}$, ${\bm \sigma}^{(2)}_{ij}$ and ${\bm \sigma}^{(4)}_{ij}$ are the same and we simply have to replace $w (|{\bm c}_l|^2)$ with $p (|{\bm c}_l|^2)$ and ${\cal G}_1$ with ${\cal G}_2$). There are, nevertheless, extra contributions coming from the energy shells with $|{\bm c}_l|^2=5,8$
\be
\begin{split}
& {\bm \sigma}^{(5)}_{ij}=\frac{{\cal G}_2}{4} \sum_{\ell=13-20} p(5) \psi({\bm x}) \psi ({\bm x}+{\bm c}_{\ell})  {\bm c}_{\ell}^i {\bm c}_{\ell}^j+  \\
&\frac{{\cal G}_2}{4} p(5) \left(\psi ({\bm x}+{\bm c}_5) \psi ({\bm x}+{\bm c}_3)+\psi ({\bm x}+{\bm c}_1) \psi ({\bm x}+{\bm c}_7) \right){\bm c}_{13}^i {\bm c}_{13}^j +\\
&\frac{{\cal G}_2}{4} p(5)\left( \psi ({\bm x}+{\bm c}_5) \psi ({\bm x}+{\bm c}_4)+\psi ({\bm x}+{\bm c}_2) \psi ({\bm x}+{\bm c}_7) \right){\bm c}_{14}^i {\bm c}_{14}^j +  \\
&\frac{{\cal G}_2}{4} p(5) \left( \psi ({\bm x}+{\bm c}_2) \psi ({\bm x}+{\bm c}_8)+\psi ({\bm x}+{\bm c}_6) \psi ({\bm x}+{\bm c}_4) \right){\bm c}_{15}^i {\bm c}_{15}^j+  \\
&\frac{{\cal G}_2}{4} p(5) \left( \psi ({\bm x}+{\bm c}_6) \psi ({\bm x}+{\bm c}_1)+\psi ({\bm x}+{\bm c}_3) \psi ({\bm x}+{\bm c}_8) \right){\bm c}_{16}^i {\bm c}_{16}^j   
\end{split}
\ee
\be
\begin{split}
{\bm \sigma}^{(8)}_{ij}=& \frac{{\cal G}_2}{4} \sum_{\ell=21-24} p(8)  \psi({\bm x}) \psi({\bm x}+{\bm c}_{\ell}){\bm c}_{\ell}^i {\bm c}_{\ell}^j+\\
& \frac{{\cal G}_2}{4} \sum_{\ell=21-24} p(8) \psi\left({\bm x}-\frac{{\bm c}_{\ell}}{2}\right)\psi\left({\bm x}+\frac{{\bm c}_{\ell}}{2}\right)   {\bm c}_{\ell}^i {\bm c}_{\ell}^j.
\end{split}
\ee
The mismatch between the normal $\sigma_{xx}$ and tangential $\sigma_{zz}$ components of the interaction pressure tensor is written as follows
$$
(\sigma_{xx}-\sigma_{zz})(x)=(\sigma^{(int)}_{xx}-\sigma^{(int)}_{zz})(x)+\frac{1}{\rho}\left(\tau-\frac{1}{2}\right)^2 {F}^2_x(x)
$$
where
\be
\begin{split}
(\sigma^{(int)}_{xx}-\sigma^{(int)}_{zz})(x)=&+A_1 \psi(x)(\psi(x+1)+\psi(x-1))\\
&+A_2 \psi(x)(\psi(x+2)+\psi(x-2)) \\
&+A_0 \psi(x)\psi(x)+A_3 \psi(x+1)\psi(x-1).
\end{split}
\ee
In the above, we have used the following definitions, directly related to the specific weights of the model
$$
A_1={\cal G}_2 \left(\frac{p(1)}{2}-3 p(5) \right)+ {\cal G}_1 \frac{w(1)}{2} 
$$
$$
A_2={\cal G}_2  \left(p(4)+\frac{3}{2}p(5) \right)+{\cal G}_1 w(4)
$$
$$
A_0={\cal G}_2 \left( -p(1)-4 p(4) \right)-{\cal G}_1 \left( w(1)-4 w(4) \right)
$$
$$
A_3={\cal G}_2 \left( 2 p(4)+3 p(5) \right)+2 {\cal G}_1 w(4).
$$
By Taylor expanding
$$
\psi(x+a) \approx \psi(x)+a \frac{d \psi(x)}{d x}+\frac{a^2}{2} \frac{d^2 \psi(x)}{d x^2 }+... 
$$
and integrating by parts (provided that $ d \rho /d x =0$ in the bulk phases at $x=\pm \infty$), one identifies the overall tension between $x=-\infty$ and $x=+\infty$ as follows:
\be
\begin{split}
\gamma& = \int_{-\infty}^{+\infty} p_s(x) dx=\int_{-\infty}^{+\infty}(\sigma^{(int)}_{xx}-\sigma^{(int)}_{zz})(x) dx= \nonumber \\
& \int_{-\infty}^{+\infty} \left({C}_2 \left(\frac{d\psi}{dx} \right)^2+{C}_4 \left(\frac{d^2 \psi}{d x^2} \right)^2+\left(\tau-\frac{1}{2}\right)^2\frac{1}{\rho(x)} F^2_x(x) \right) dx
\end{split}
\ee
where
\be
\begin{split}
& C_2=-A_1-4 A_2-2 A_3= \nonumber \\
&-{\cal G}_1\left(\frac{w(1)}{2}+8 w(4) \right)-{\cal G}_2\left(\frac{p(1)}{2}+8 p(4)+9 p(5)\right)
\end{split}
\ee
\be
\begin{split}
&C_4=\frac{1}{12}A_1+\frac{4}{3} A_2+\frac{2}{3} A_3= \nonumber \\ 
&+{\cal G}_1\left(\frac{w(1)}{24}+\frac{8}{3}w(4) \right)+{\cal G}_2\left(\frac{p(1)}{24}+\frac{8}{3}p(4)+\frac{15}{4}p(5) \right).  
\end{split}
\ee

\section{Appendix B}
The second order isotropy is given by
$$
\sum_l W(|{\bm c}_l|) {\bm c}_l^{i_1}{\bm c}_l^{i_2}=\delta_{i_1 i_2}.
$$
The forth order isotropy is given by
$$
\sum_l W(|{\bm c}_l|^2) {\bm c}_l^{i_1}{\bm c}_l^{i_2}{\bm c}_l^{i_3}{\bm c}_l^{i_4}=e_4(W)\left(\delta_{i_{1}i_{2}}\delta_{i_{3}i_{4}}+\delta_{i_{1}i_{3}}\delta_{i_{2}i_{4}}+\delta_{i_{1}i_{4}}\delta_{i_{2}i_{3}}\right).
$$
As for the sixth order isotropy, we find
$$
\sum_l W(|{\bm c}_l|^2) {\bm c}_l^{i_1}{\bm c}_l^{i_2}{\bm c}_l^{i_3}{\bm c}_l^{i_4}{\bm c}_l^{i_5}{\bm c}_l^{i_6}=e_6(W)\left(\delta_{i_{1}i_{2}}\delta_{i_{3}i_{4}}\delta_{i_{5}i_{6}}+ ...\right)
$$
where $(+...)$ accounts for all the cyclyc permutations. From the above expressions, we easily obtain (we stop at the shell $|{\bm c}_l|^2=8$)
$$
\sum_l W(|{\bm c}_l|^2) ({\bm c}_l^{x})^{2n}({\bm c}_l^{z})^{2m}=e_{2n+2m}(W) (2m-1)!!(2n-1)!!
$$
and, hence
\be\label{e4part}
\begin{split}
e_4(W) = &\sum_l W(|{\bm c}_l|^2) ({\bm c}_l^{x})^{2}({\bm c}_l^{z})^{2}=\\
& 4 W(2)+32 W(5)+64 W(8)
\end{split}
\ee
\be\label{e6part}
\begin{split}
3 e_6(W) = &\sum_l W(|{\bm c}_l|^{2}) ({\bm c}_l^{x})^{4}({\bm c}_l^{z})^{2}=\\
& 4 W(2)+80 W(5)+256 W(8).
\end{split}
\ee
By varying the indexes $n$, $m$ one easily obtains a set of independent constraints, which are equivalent to (details are fully reported in \cite{Shan06,Sbragaglia07})
\be\label{4sum}
W(1)-4 W(2)+16 W(4)-14 W(5) -64 W(8)=0
\ee
\be\label{6sum}
W(1)-8 W(2)+64 W(4)-70 W(5)-512 W(8)=0
\ee
Combining (\ref{e4part}-\ref{e6part}) with (\ref{4sum}-\ref{6sum}), we get
\be\label{finale4}
e_4(W)= W(1)+16 W(4) +18 W(5)
\ee
\be\label{finale6}
e_6(W)=\frac{W(1)}{6}+\frac{32}{3} W(4)+15 W(5)
\ee
which are reported and used in the text. We note that the expression for $e_4(W)$ reported in (\ref{finale4}), due to the property (\ref{4sum}),  can be recasted also in the form 
\be
\begin{split}
e_4(W)=& W(1)+16 W(4) +18 W(5)=\\
&\frac{W(1)}{2}+\frac{W(1)}{2}+16 W(4) +18 W(5)=\\
&\frac{W(1)}{2}+2 W(2)+8 W(4)+25 W(5)+32 W(8) 
\end{split}
\ee
that coincides with the expression reported in the paper by Shan \cite{Shan08} (see equations before equation (10)).

\bibliography{rsc}       
\bibliographystyle{rsc} 
\end{document}